\documentclass[%
reprint,
twocolumn,
groupedaddress,
amsmath,amssymb,
prl,
]{revtex4-1}
\usepackage{graphicx}
\usepackage{hhline}
\usepackage{multirow}
\usepackage{color}
\usepackage{times}
\usepackage{url}
\usepackage{tikz}
\usepackage{tikz-feynman}
\usepackage{dsfont}
\usepackage{mathtools}
\usepackage{dcolumn}
\usepackage{setspace}
\usepackage{ulem}
\usepackage{physics}
\usepackage{comment}
\usepackage{hyperref}
\usepackage{bbm}
\usepackage[bottom]{footmisc}

\usepackage[toc,page,titletoc]{appendix}
\usepackage{amsfonts}

\begin{document}
	\title{Non-Abelian charge conversion in bilayer binary honeycomb lattice systems}
	
	\author{Chiranjit Mondal}
        \thanks{These authors contributed equally}	
        \author{Rasoul Ghadimi}
        \thanks{These authors contributed equally}
	\author{Bohm-Jung Yang}
	\email{bjyang@snu.ac.kr}

	\affiliation{Department of Physics and Astronomy, Seoul National University, Seoul 08826, Korea}
	\affiliation{Center for Theoretical Physics (CTP), Seoul National University, Seoul 08826, Korea}
	\affiliation{Institute of Applied Physics, Seoul National University, Seoul 08826, Korea}
	
	\date{\today}

\begin{abstract}
In two-dimensional systems with space-time inversion symmetry, Dirac nodes (DNs) carry non-Abelian topological charges which induce intriguing momentum space braiding phenomenon. Although the original idea was proposed in condensed matter setup, the experimental verification of non-Abelian charge conversion has been limited to artificial metamaterials because of the difficulty in identifying suitable materials in which controlled tuning of DN positions is possible. In this work, we propose bilayer binary honeycomb lattices (BBHL) as a new material platform to study the non-Abelian charge conversion phenomenon in which DN positions in momentum space can be manipulated. More explicitly, we demonstrate that layer sliding and vertical pressure serve as tunable braiding parameters controlling the non-Abelian charge conversion process which is crucial to understand the stacking-dependent electronic properties of BBHL systems. We show that the BBHL systems are a promising candidate for the experimental realization of non-Abelian phenomena of DNs in condensed matter.
	\end{abstract}
	\date{\today}
	 \maketitle

Recently, the braiding properties of Dirac nodes (DNs) in two-dimensional (2D) systems,  have garnered considerable attention ~\cite{PhysRevX.9.021013, doi:10.1126/science.aau8740, Bouhon2020,doi:10.1126/science.adf9621, Guo2021, Peng2022,PhysRevLett.125.053601,lee2024euler}.  
In these systems, the DNs are protected by space-time inversion symmetry ($I_\text{ST}$) where $I_\text{ST}$ represents either the combination of space-inversion ($\mathcal{P}$) and time reversal ($\mathcal{T}$) i.e. $\mathcal{PT}$ for spinless systems or combination of the two-fold rotation ($ C_{2}$) symmetry and $\mathcal{T}$ i.e. $ C_{2}\mathcal{T}$ for both the spinless and spinfull systems ~\cite{ PhysRevB.89.235127, PhysRevB.92.081201, PhysRevB.91.161105, PhysRevLett.118.056401, PhysRevLett.118.156401, PhysRevLett.119.226801, PhysRevLett.121.106403, PhysRevB.99.235125, Ahn_2019}.
A pair of bands in $I_\text{ST}$ symmetric 2D systems can have an integer topological invariant, called the Euler number. 
Two bands with nonzero Euler number carry several DNs in between such that the total vorticity of DNs is equal to two times of the Euler number. 
Thus, two DNs associated with nonzero Euler number cannot be annihilated by merging although their total Berry phase is zero modulo $2\pi$ \cite{PhysRevX.9.021013}. 
Interestingly, the two DNs associated with nonzero Euler number can be annihilated through braiding around the adjacent band's DNs, which is neatly formulated in terms of DN vorticity reversal by crossing Dirac strings \cite{PhysRevX.9.021013} or non-Abelian topological charge conversion~\cite{Bouhon2020,doi:10.1126/science.aau8740,doi:10.1126/science.adf9621}.
It is suggested that such braiding can be realized in a three-dimensional (3D) ZrTe \cite{Bouhon2020}, the phononic band structure of  2D layered silicate materials \cite{Peng2022}, 2D magnetic materials with in-plane magnetization \cite{lee2024euler}, optical lattices \cite{jankowski2023optical}, photonic system \cite{PhysRevLett.129.263604,PhysRevLett.130.083601,Sun2022,Noh2020,PhysRevLett.125.033901}, classical systems including acoustic and electric circuits \cite{Chen2022,Qiu2023,PhysRevLett.124.146801,Jiang2021,PhysRevB.105.214108,PhysRevLett.128.246601,Wu2022,Jiang2021,PhysRevB.103.205303} and transmission line networks \cite{Guo2021}.
Nevertheless, despite the theoretical proposals of the conversion phenomena of DN non-Abelian charges, a clean and tunable 2D electronic material platform is still required for experimental exploration in condensed matter systems.
Such a platform would give a new route to explore, confirm, and understand non-Abelian braiding phenomena of DNs.

In this letter, we show that bilayer binary honeycomb lattice (BBHL) systems, such as bilayer h-BX (where X = N, P, or As), exhibit  $I_\text{ST} = \mathcal{PT}$ symmetry that offers clean and tunable platforms to explore multi-gap topology and its non-Abelian braiding phenomena.
Using tight-binding calculation and \textit{ab-initio} simulations, we show that the braiding of DNs in these setups can easily be manipulated by external stimuli such as layer-sliding, pressure, and sublattice potential imbalances.  
The layer-sliding naturally appears in BBHL moir\'e systems, thus presenting them as a promising avenue for exploring multigap topology and non-Abelian charge conversion physics. 

\begin{figure}[t!]
	\centering
	\includegraphics[width= \linewidth]{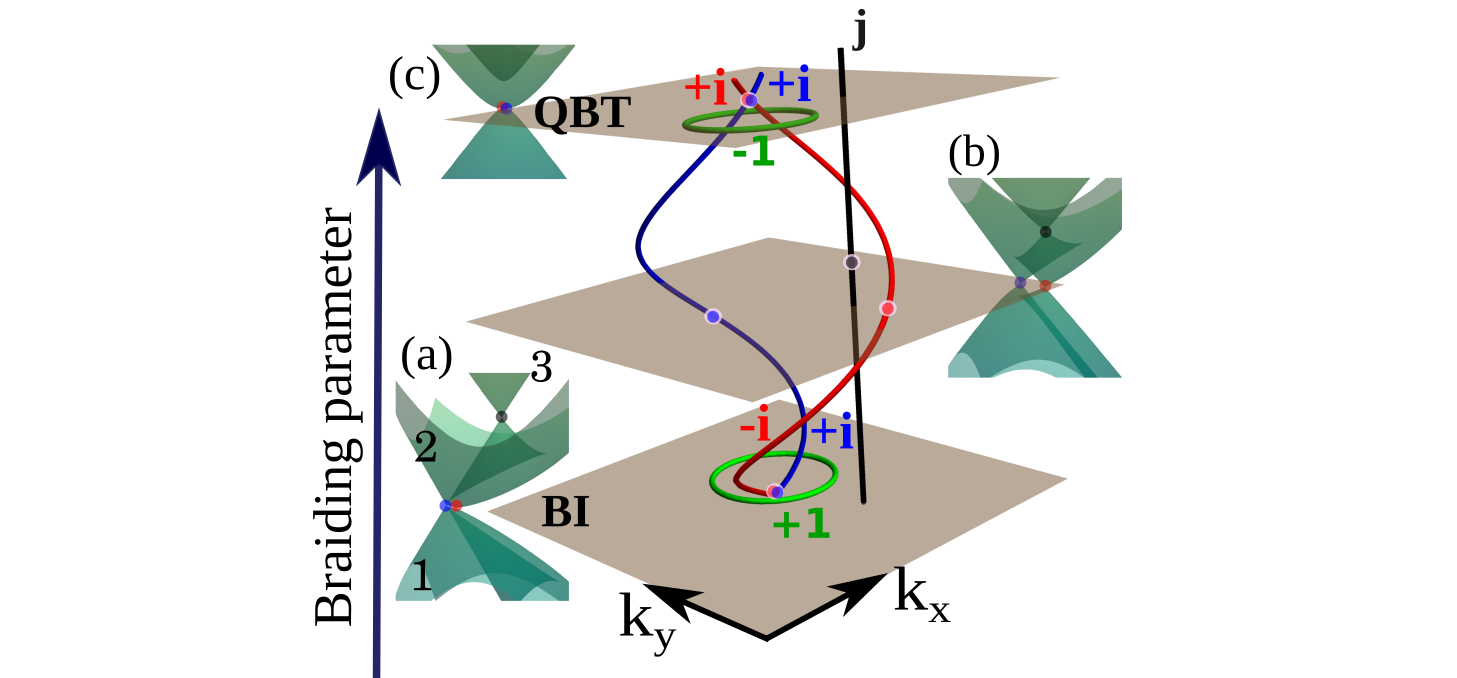}
\caption{
 Schematic representation of Dirac node (DN) braiding processes and the related charge conversion protocol in low energy three bands of honeycomb bilayer systems.
(a) Two DNs with opposite charges ($i$, $-i$) are created by the band inversion (BI) between bands 1 and 2.
Another DN exists between bands 2 and 3 with charge $j$.
(b,c) One DN with charge $-i$ (indicated by red) encircles the DN with charge $j$ (indicated by black) during the braiding parameter evolution and converts its charge from $-i$ to $+i$ and eventually meets with the blue DN with charge $+i$ and form a quadratic band touching (QBT) with total charge $-1$.
The green loop represents the charge-calculating loop enclosing the red and blue DNs. $+1$ ($-1$) represents the trivial (non-trivial) charge on the loops. }
	\label{fig1}
\end{figure}

\textit{Non Abelian charge of DNs}-.
Let us first briefly recap the idea of DN charge conversion in 2D systems with $I_\text{ST}$ symmetry ~\cite{PhysRevX.9.021013,Bouhon2020,doi:10.1126/science.aau8740}.
In a suitable basis where $I_\text{ST} = \mathcal{K}$ ($\mathcal{K}$ is a complex conjugation operator), the Hamiltonian of the system becomes real i.e., $H(k)=H^*(k)$ \cite{PhysRevX.9.021013}.
For such Hamiltonian with more than two energy bands, the underlying symmetry group has a non-Abelian group structure. 
For instance, a three-band Hamiltonian has the generic form of $H(k) = \sum_{n=1}^{3} \epsilon_n \ket{\psi_k^n}\bra{\psi_k^n}$, where $\ket{\psi_k^n}$ is a real orthonormal three component vector with $O(1)$ ambiguity with the eigenvalue $\epsilon_n$. 
Therefore, the corresponding classifying space is $M_3 = O(3)/O(1)^3$ where $O(N)$ represents the orthogonal group. 
The homotopy group of the system is quaternion i.e, $\pi_1(M_3) = \mathbb{Q}$, a non-Abelian group which has five conjugacy classes $\{+1, -1, \pm i, \pm j, \pm ij\}$ with the anti-commuting relations $ij=-ji$ and $i^2 = j^2 =-1$ \cite{doi:10.1126/science.aau8740}. 
In the $I_\text{ST}$ symmetric three bands labeled by 1, 2, and 3 in increasing energy order, $\pm i$ ($\pm j$) represents the charge of nodes between bands 1 and 2  (2 and 3), where $\pm$ determine relative charges of the nodes within the same pair of bands.
$+1$ indicates that no node exists or that all nodes can be annihilated pairwise, while $-1$ indicates a quadratic band touching or at least two DNs with the same charge \cite{doi:10.1126/science.aau8740}. 
In Fig.~\ref{fig1}, we schematically show how two DNs with opposite charges ($\pm i$) are created by a band inversion [Fig.~\ref{fig1} (a)] and converted to a quadratic band touching (QBT) [Fig.~\ref{fig1} (c)]. 
In this process, one of the two DNs changes its sign by encircling a DN of adjacent bands with charge $j$ [Fig.~\ref{fig1} (b)]. 
This charge conversion process is called non-Abelian braiding of DNs. 
Therefore, the $I_\text{ST}$ symmetric band nodes are characterized by the non-Abelian quarternion charge, which serves as the key ingredient for the non-Abelian braiding protocol and related charge conversion phenomena \cite{doi:10.1126/science.aau8740, Bouhon2020} [see Supplemental Material (SM) \footnote{see Supplemental Material in ***  which includes Refs. \cite{Bouhon2020,doi:10.1126/science.aau8740,PhysRevX.9.021013,RevModPhys.51.591,PhysRevB.50.17953,PhysRevB.47.558,PhysRevB.59.1758,PhysRevB.56.12847,PhysRevB.65.035109,RevModPhys.84.1419,MOSTOFI20142309,TramblydeLaissardière2010,PhysRevB.87.205404}.} for detailed discussions on non-Abelian charges].  

The relative charges of two DNs within the same bands can also be determined using the patch Euler class ($\chi$) \cite{Bouhon2020}. 
The patch Euler class over a patch $\mathcal{D}$, for two real sub-bands $\ket{u^1(k)}$, $\ket{u^2(k)}$ that are separated by a gap from the adjacent bands in the given patch $\mathcal{D}$ is,~\cite{Bouhon2020},
\begin{equation}\label{equ:eulerclass}
    \chi(\mathcal{D})=\tfrac{1}{2\pi}\left[\int_{\mathcal{D}} \text{Eu}(\mathbf{k}) dk_xdk_y-\oint_{\partial \mathcal{D}} \mathbf{a}(\mathbf{k})\cdot d\mathbf{k}\right],
\end{equation}
where Euler connection and Euler form are given by $\mathbf{a}(\mathbf{k})=\braket{u^1(k)}{\nabla_{\mathbf{k}} u^2(k)}$, and $\text{Eu}(\mathbf{k})=\bra{\nabla_{\mathbf{k}} u^1(\mathbf{k})}\times \ket{\nabla_{\mathbf{k}} u^2(\mathbf{k})}$, respectively \cite{Bouhon2020,PhysRevX.9.021013}. 
The patch Euler class for a single DN is $\chi= \pm \frac{1}{2}$, where its sign depends on their charges and making it $\chi=\tfrac{1}{2}+\tfrac{1}{2}=1$ (non-trivial) or $\chi=\tfrac{1}{2}-\tfrac{1}{2}=0$ (trivial) when two DNs with same or opposite charges reside inside the patch.
In the following, we use the patch Euler class to discuss braiding phenomena in the BBHL system where layer-shifting/moir\'e-potential or pressure can be employed as the tunable braiding parameters.

\begin{figure}[t!]
	\centering
	\includegraphics[width=\linewidth]{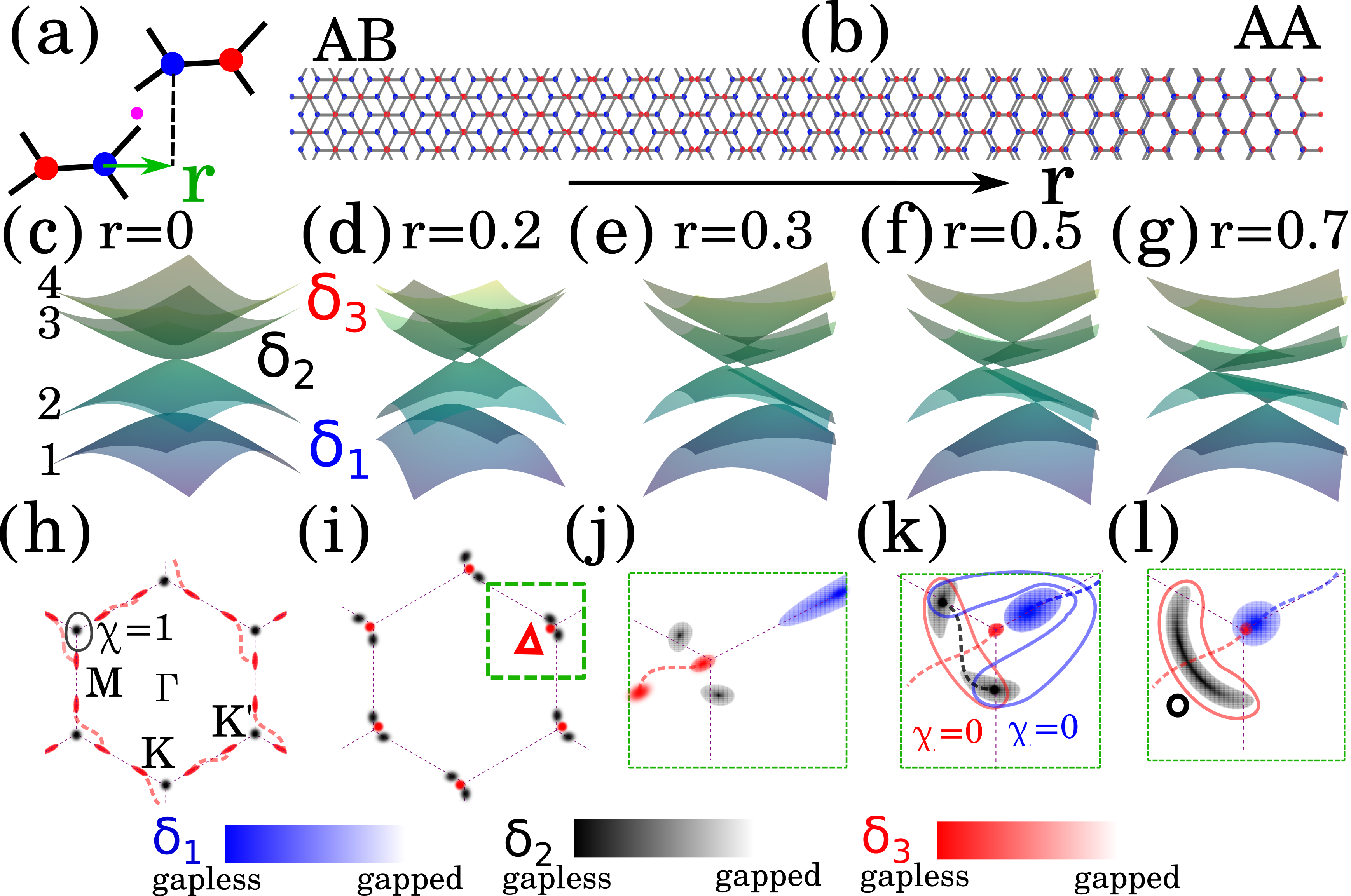}
\caption{ 
Layer sliding as a braiding parameter.
(a) Hexagonal bilayer with two sublattices indicated by blue and red dots.
The different stacking configurations can be generated by shifting one layer with respect to the other layer (indicated by green arrow $r$). 
The pink dot in (a) represents the inversion centers that connect the two same-colored sites in the top and bottom layers, respectively. 
(b) A snap shot of moir\'e of hexagonal bilayer that connects AB ($r=0$) to AA ($r=1$) stacking. 
For large moir\'e length, a local position on (b) can be considered as a bilayer system with a certain amount of shifting.
(c-g) Energy dispersion (around one valley) for various $r$ and three gaps represented by $\delta_1$, $\delta_2$, and $\delta_3$ between the band 1,2,3, and 4.
(h,i) The distribution of the gap size $|\delta_{1,2,3} |$ in the Brillouin zone
(j-l) $|\delta|$ distribution near K valley.
The dotted lines in (h-l) are Dirac strings, which can flip the topological charge of the adjacent DN after crossing.  
$\bigcirc$ and $\triangle$ represent the regions where creation and annihilation of DNs occurs, respectively. 
The blue, black, and red color bars at the bottom indicate the gapless (dark color) and gapped (light color) regions in $\delta_1$,$\delta_2$, and $\delta_2$ gap respectively, in the momentum space.} 
	\label{fig2}
\end{figure}

\textit{Layer sliding as a braiding parameter-.}
Our model is motivated by the energy band structure of bilayer graphene (BLG) with $I_{\text{ST}}$ symmetry. 
It is known that BLG has two highly symmetric stacking configurations, AB and AA stacking \cite{doi:10.1021/jp511692e}.
AB stacking provides the well-known QBT at the Brillouin zone corner $K$ and $K'$ whereas AA stacking gives a nodal line at $K$ and $K'$
\cite{RevModPhys.81.109,Zhang2009,PhysRevB.86.075439,PhysRevB.106.L121118}.  
These stacking can be smoothly transformed to each other by layer sliding in $I_{\text{ST}}=\mathcal{PT}$  symmetric manner as shown in Fig.~\ref{fig2}(a). 
The inversion center is indicated by the pink circle in Fig.~\ref{fig2}(a) which connects the same sublattice in the top and bottom layers. 
The green arrow represents the direction of layer sliding.  
If $r=0$ ($r=1$), the stacking becomes identical to the AB (AA) configurations.
An arbitrary value of $r$ corresponds to a stacking between AB and AA configurations which can be achieved by layer sliding.
All such stacking configurations can also naturally appear in suitable moir\'e lattice [see Fig.~\ref{fig2}(b)], where locally atomic stacking resembles two graphene layers with a certain amount of sliding $r$. 
Additionally, different stacking of BLG and its moir\'e structures already has been constructed in metamaterials providing further tunable setup to study our proposed braiding phenomena \cite{PhysRevLett.120.116802, PhysRevLett.125.214301,PhysRevB.102.180304, PhysRevB.101.121103}.

Fig.~\ref{fig2}(c-g) shows the evolution of energy dispersion around $K$ with the corresponding nodal structures in (Fig.~\ref{fig2}(h-l)) upon increasing $r$, which shows that the topologically protected QBT at $r=0$ splits into two DNs that migrate in momentum space until they are pair-annihilated as $r$ increases. 
We use small sublattice potential which respects $I_{\text{ST}}=\mathcal{PT}$, to complete the braiding process as AA stacking without sublattice mass results in nodal loops \cite{PhysRevB.106.L121118}.
This sublattice mass can be generated, for instance, by electronic instabilities \cite{PhysRevLett.109.206801,PhysRevB.87.121401} or mimicking systems such as bi-layers h-BX (X= P and As) \cite{ghadimi2024quantum}.
As each layer has two sublattices, we obtain four energy bands (denoted by 1,2,3, and 4 in Fig.~\ref{fig2}(c)). 
Three energy gaps between the bands are 
$\delta_1$ [between 1 and 2], 
$\delta_2$ [between 2 and 3], and $\delta_3$ [between 3 and 4]  
[see SM for the details of the \textit{tight-binding} and \textit{ab-initio} calculations].

Fig.~\ref{fig2}(h-l) shows the gap values of $ |\delta_{1,2,3}| $, represented by blue, black, and red colors in the three gaps. 
The dark blue, black, and red color indicates the gapless nodal points in the $\delta_1$, $\delta_2$, and $\delta_3$, respectively.
Fig.~\ref{fig2}(h) shows that the energy dispersion contains DNs near the M (red nodes) point in the $\delta_1$ gap and the QBT (black nodes) in the $\delta_2$ gap at $K/K'$. 
We obtain that $\chi$ of the QBT for the region indicated by the black circle around $K$ in Fig.~\ref{fig2}(h) gives $\chi=1$, and confirms that the QBT is topologically protected.
At $r=0.2$, the QBT splits into two DNs (black nodes) [see Fig.~\ref{fig2}(d,i)].
Also, at the same time, some of the DNs undergo pair annihilation and pair creation in other gaps. 
The region where pair-annihilation (pair-creation) occurs is indicated by the ``O" (``$\Delta$") symbol. 
At $r=0.2$, a pair of DNs (red nodes) is generated in the $\delta_3$ gap as shown in Fig.~\ref{fig2}(d, i) around $K$ and $K'$. 
Fig.~\ref{fig2} (j) shows the magnified view of the nodal structure around $K$ for $r=0.3$ which shows one red node remains near $K/K'$ while the other moves towards the center of BZ.
At $r\approx 0.5$, one DN in $\delta_1$ gap (blue nodes) moves towards the $K$ [see Fig.~\ref{fig2}(f, k)].
In this situation, two black nodes in the $\delta_2$ gap stay with the largest separation in momentum space.
We compute the $\chi$ of these two black DNs in two patches shown by blue and red loops in  Fig.~\ref{fig2}(k), both of which give a trivial Euler class (i.e. $\chi=0$).
Therefore the two black nodes on the red or blue patch have opposite charges and can undergo pair-annihilation without any topological obstruction [Fig.~\ref{fig2}(g,l)].

To intuitively understand the role of braiding protocol, we connect the DN pairs (within the same gap) using dashed lines to represent their  Dirac strings \cite{PhysRevX.9.021013}.  
As mentioned earlier, a DN flips its sign when it encircles adjacent band DNs. 
Dirac strings provide a useful tool to determine whether a Dirac cone has undergone such an encircling process \cite{PhysRevX.9.021013}. 
Intuitively, a Dirac string arises as a consequence of the $\pi$ Berry phase associated with the DN, where the wave function in the real basis exhibits a phase discontinuity when smoothly sweeping the loop that encloses a DN.
If two DNs initially have the same charges, after one of them crosses a Dirac string between DNs in adjacent gaps, they become to carry the opposite charges as shown in Fig.~\ref{fig2}(h-l).
Accordingly, with increasing $r$, two black DNs with the same charges eventually become oppositely charged when they move inside the red patch crossing the red Dirac string, and thereby become pair-annihilated after the collision as shown Fig.~\ref{fig2}(k-l). 

In summary, the two black DNs, which initially carry the same charges ($+i,+i$), turn into carrying the opposite charge ($+i,-i$) after one of them encircles a red DN, thus eventually becomes pair-annihilated.
Note that flipping the sign of sublattice potential $m\rightarrow -m$ mediates braiding with the annihilation process inside the blue patch in Fig.~\ref{fig2}(k) (discussed in SM). 
In the case of BLG with m=0, the braiding process is not completed, even when $r$ varies from $0$ to $1$, because the final AA-stacked state with $r=1$ possesses a nodal line at the Fermi level protected by mirror symmetry ($\mathcal{M}_z: z\rightarrow -z$) \cite{PhysRevB.86.075439,PhysRevB.106.L121118}.

\begin{figure}[t!]
	\centering
	\includegraphics[width=1\linewidth]{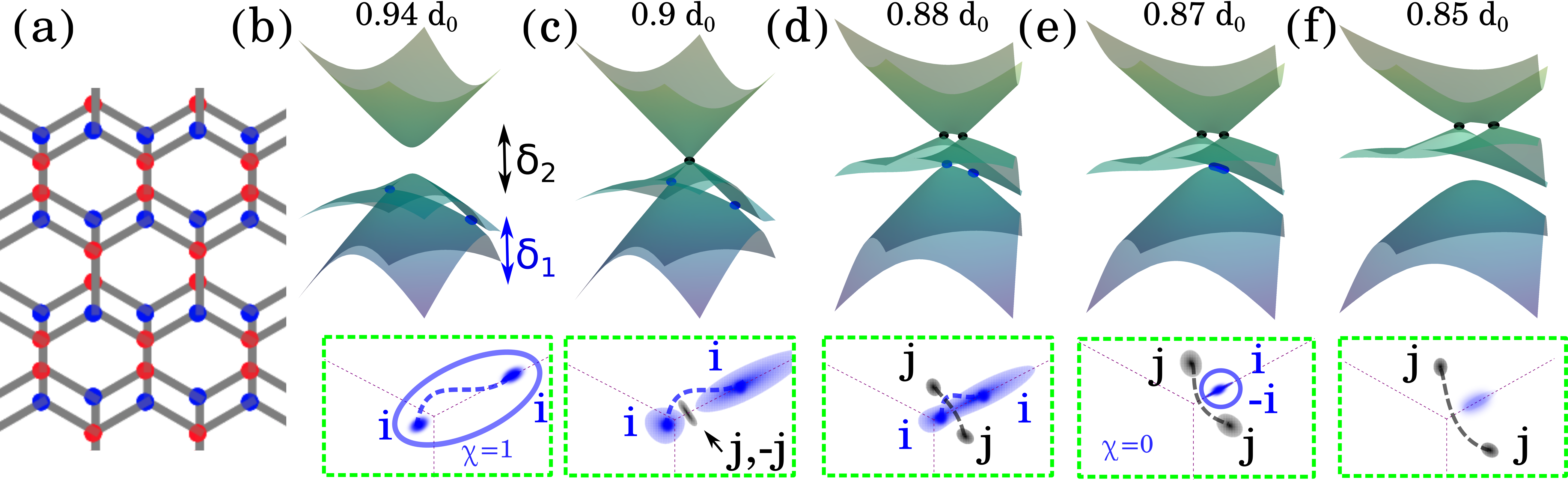}
	\caption{Pressure-induced braiding.
 (a) A bilayer honeycomb lattice with a small shift ($r=0.1$) of one layer from AB configuration. 
 (b-f) The dispersion energy (upper panel) and configuration of the gapless region in momentum space near K point (lower panels) are plotted with  decreasing interlayer distance (increasing pressure). $d_0$ is the interlayer distance of BLG. $\chi$ becomes non-trivial to trivial during the evolution.
 }
	\label{fig3}
\end{figure}

\textit{Pressure as a braiding parameter-.}
Let us propose another braiding protocol in BBHL induced by pressure. 
We find that under external pressure, a braiding mechanism mediates the transportation of DNs from one gap to the other gap. 
To simulate the pressured BBHL, we decrease the interlayer distance in both the \textit{tight-binding} model and DFT calculations.

In the following, we show the non-Abelian charge conversion using the \textit{tight-binding} model. 
(see SM for the DFT results). 
Fig.~\ref{fig3}(a) shows BBHL stacking with $r=0.1$.
We consider the three lower bands and two gaps $\delta_1$ and  $\delta_2$ between the bands.
For large sublattice potential $m=-0.4 eV$ (mimicking h-BX), two DNs with the same charge ($i$,$i$) are initially located in the $\delta_1$ gap [Fig.~\ref{fig3}(b)].  
At the interlayer distance $0.9 d_0$ where $d_0$ indicates the interlayer distance of pristine bilayer graphene [Fig.~\ref{fig3}(c)], a band touching happens in the $\delta_2$ gap which creates a pair of DNs in $\delta_2$ gap (black nodes) with opposite charges ($j$,$-j$) though the BI. 
Further reduction of interlayer distance [Fig.~\ref{fig3}(d)] separates black DNs and one of them crosses the blue Dirac string.
Therefore the mutual charges of black nodes are converted from ($j$,$-j$) to ($j$,$j$). 
With additional pressure [Fig.~\ref{fig3}(e)]  the blue nodes cross the black Dirac strings and their charges are converted from ($i$,$i$) into ($-i$,$i$).
As such the blue nodes in $\delta_1$ gap are annihilated after mutual collision [Fig.~\ref{fig3}(e,f)]. 
The transportation of DNs from $\delta_1$ gap to $\delta_2$ gap occurs by the non-Abelian braiding and a sequence of charge conversion processes. In SM, we discuss the charge conversion process for various $m$ and interlayer distance/pressure which exhibit similar phenomena. 
For the AB stacking (i.e. $r=0$), the two DNs combine to form a QBT because of the underlying $\mathcal{C}_{3z}$ rotation symmetry, making the visualization of the Dirac string difficult. Nonetheless, the transportation of the QBT from one gap to another gap occurs by the non-Abelian braiding when the external pressure is tuned (see SM). 
However, the nodal transition always passes through a triple point degeneracy (TDP) which makes $\chi$ ill-defined at the TDP, necessitating the frame rotation charge calculation \cite{Bouhon2020} (see SM for the detail). 
We also note that the QBT can be unstable under trigonal wrapping. However, it does not affect the braiding phenomena (see SM).

\begin{figure}[t!]
	\centering
	\includegraphics[width=0.9\linewidth]{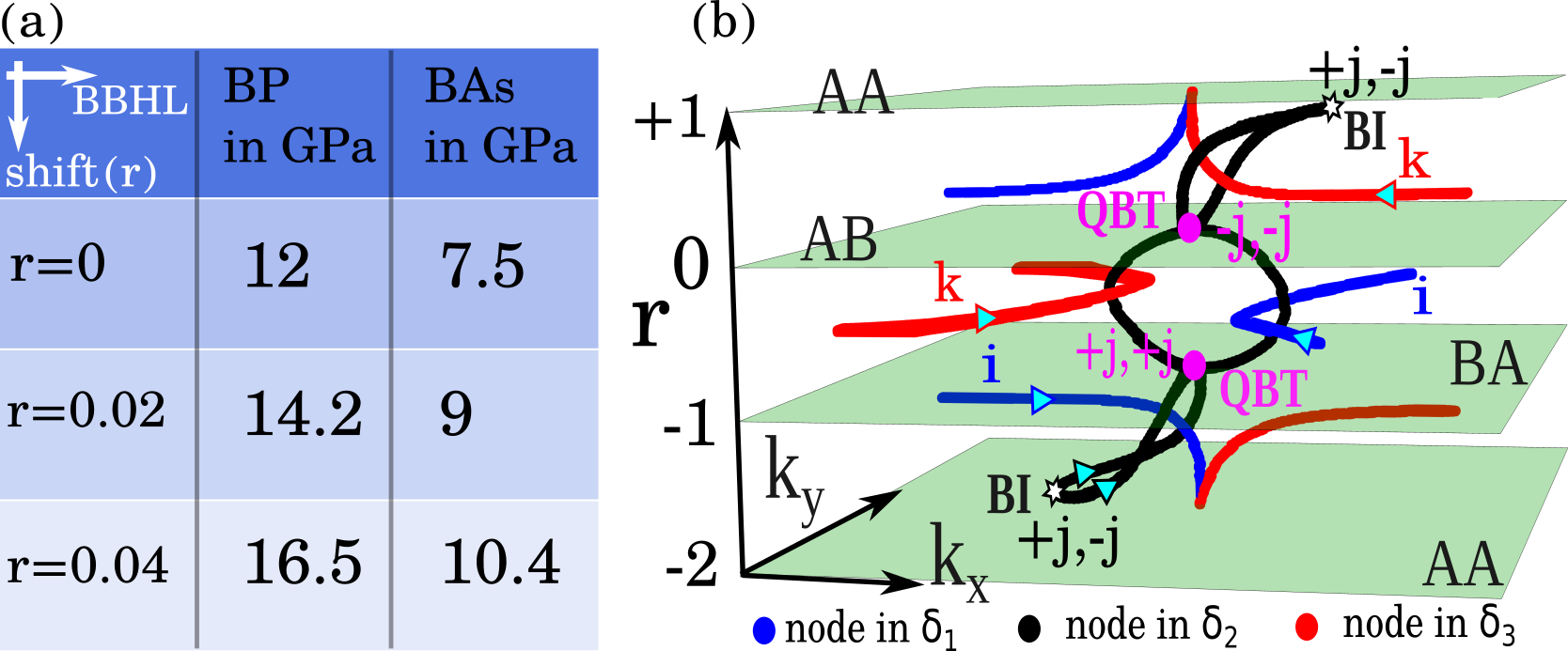}
	 \caption{ (a) Maximum pressure in gigapascal (GPa) which is required to complete the braiding process in the BBHL systems BP and BAs for the specific $r$.
(b)Summary of nodal trajectories evolution.
 Colored lines represent the evolution of Dirac nodes (DNs) in the gaps $\delta_1$ [blue], $\delta_2$ [black], and $\delta_3$ [red], shown in Fig.~\ref{fig2}(c), as a function of layer shifting ($r$) in momentum space near the K/K' points. 
 The green planes indicate the AA, AB, and BA stacking configurations. 
 Two black DNs with opposite charges are created by band inversion (BI) near the AA stacking (at $r=-2$ and $r=+1$) and undergo braiding with adjacent bands nodes (blue or red), eventually transforming into a quadratic band touching (QBT).
 The QBT from the BA stacking ($r=-1$) is braided by both the blue and red nodes, but its charges remain unchanged, resulting in QBT again at the AB stacking ($r=0$). The magenta arrows represent the direction of movement of DNs when $r$ is changed from $-2$ to $+1$.
 }
\label{fig4}
\end{figure}

Our DFT calculations confirm that the maximum pressure of approximately up to 12 GPa (7.5 GPa) is required in h-BP (h-BAs) for a complete braiding process (see Fig.~\ref{fig4}(a) and SM for further detail). 
To apply such pressure, we decrease the interlayer distance of h-BP (h-BAs) to $0.81 d_0$ ($0.85 d_0$ ) where $d_0 = 3.69 \AA (3.75 \AA)$ is the relaxed interlayer distance for h-BP (h-BAs). 
It is important to mention that such pressure has been experimentally applied in graphite-like h-BN system where a similar amount of interlayer compression is observed \cite{SOLOZHENKO19951}. Also, recent experiment demonstrated the measurement of the dynamical band structure in moir\'e system under external pressure \cite{Yankowitz2018}.
In Fig.~\ref{fig4}(a), we show the maximum pressure value (calculated using DFT) required for a specific $r$ to complete the braiding process in the two representative BBHL systems BP and BAs.

\textit{Nodal trajectories during the braiding-.}
As we discussed earlier, DNs in $I_\text{ST}$ systems are topologically protected and can only be annihilated in pairs. 
By tracking the positions of these DNs in momentum space, one can obtain continuous trajectories as functions of the braiding parameters.
These trajectories often are intertwined with trajectories of other DNs from other bands and provide complete information on the braiding processes.
 In Fig.~\ref{fig4}(b), we show the nodal trajectory in momentum space (horizontal plan) around the K/K' point where layer shifting ($r$) acts as the braiding parameter (vertical axis). 
The layer stacking changes sequentially from AA ($r=-2$) to AB ($r=-1$), then to BA ($r=0$), and finally returns to AA ($r=+1$). 
Near the AA stacking ($r=-2$), a BI creates two black nodes with opposite charges [we assume small $m$; see SM for different $m$ and pressure]. 
 They go through a braiding process crossing the Dirac strings between the blue nodes, which makes them have the same charge, leading to QBT in BA stacking ($r=-1$). 
 Then the QBT at $r=-1$ again splits into two DNs which undergo two braiding processes with DNs in both lower and upper gaps.
 This process makes the two black DNs to have the same charge, which eventually gives another QBT in AB stacking ($r=0$).
 Further increasing $r$, the QBT of AB stacking splits into two DNs which initially have the same charge but be converted to have the opposite charges after braiding around the red DN, thus eventually be pair-annihilated near AA stacking.
We note that when m=0, the black trajectory at AA stacking evolves into a circle, which is a consequence of the nodal line in AA stacking.

\textit{Discussion-.}
Experimentally, the evolution of energy dispersion can be observed using current techniques such as angle-resolved photoemission spectroscopy (ARPES) \cite{PhysRevLett.103.226803,Cheng2015} or locally using nano-ARPES \cite{Avila2013,Johansson2014}. 
Additionally, recently several proposals have been given to directly measure the charge of nodes \cite{doi:10.1126/science.abm6442} using interferometry \cite{breach2024interferometry} or optical responses \cite{jankowski2023optical}. 
The sliding of graphene nanoflakes on another layer of graphene has been studied using experimental techniques such as the frictional force microscope, and scanning tunneling microscopy (STM) \cite{doi:10.1021/nn305722d}. 
Our proposed theory can also be studied in a meta-materials platform such as acoustic twisted bilayer graphene or photonic systems where BBHL type electronic structures were explored ~\cite{PhysRevB.103.064304,Gardezi_2021,PhysRevB.103.214311,PhysRevB.102.180304,PhysRevB.103.214311,PhysRevApplied.17.034061}.

\begin{acknowledgements}
 C.M., R.G, and B.J.Y. were supported by Samsung Science and Technology Foundation under Project No.  SSTF-BA2002-06, National Research Foundation of Korea  (NRF) grants funded by the government of Korea (MSIT)  (Grants No. NRF-2021R1A5A1032996), and GRDC(Global  Research Development Center) Cooperative Hub Program through the National Research Foundation of Korea(NRF)  funded by the Ministry of Science and ICT(MSIT) (RS-2023 00258359)).
\end{acknowledgements}

 \bibliography{refs}
	\end{document}